\documentstyle[12pt]{article}
\begin{document}
\begin{center}
{\Large \bf  Generalized Coherent States and Spin $S\geq 1$ Systems}\\
\vspace{1.5cm}
{\large  V.G. Makhankov$^a$, M. Ag\"uero Granados$^b$, and A.V.
Makhankov$^a$} \\ {\em $^a$  Center for Nonlinear Studies, Los Alamos
National Laboratory, P.O.Box 1663, Los Alamos, N.M. 87545, USA  \\ $^b$
Universidad Aut\'onoma de Zacatecas, Centro de Estudios
Multidisciplinarios, Apartado Postal 597-C, 98068 Zacatecas, Zac. M\'exico.} \\
\vspace{1.0cm}
\end{center}
\begin{abstract}
\item   Generalized Coherent States (GCS) are constructed (and discussed)
in order to study quasiclassical behaviour of quantum spin models of the
Heisenberg type. Several such models are taken to their semiclassical limits,
whose form depends on the spin value as well as the Hamiltonian symmetry.
In the continuum approximation, SU(2)/U(1) GCS when applied give rise to the
well-known Landau-Lifshitz classical phenomenology.
For arbitrary spin values one obtains a lattice of coupled nonlinear
oscillators. Corresponding classical continuum models are described as 
well. \end{abstract}
\vspace{1.5cm}
\begin{enumerate}

\item In what follows our main aim is to provide a semiclassical
description of quantum spin models of the Heisenberg type. For the sake of
simplicity we confine ourselves to considering Heisenberg ferromagnets
(and antiferromagnets) with uniaxial anisotropy of the following form:

\begin{equation} {\hat H}_e =\mp J\sum_n (\hat {\vec S}_n\hat {\vec
S}_{n+1} +\delta {\hat S}_n^z{\hat S}^z_{n+1}) \end{equation}

(exchange anisotropy) or
\begin{equation}
{\hat H}_s =\mp J\sum_n (\hat {\vec S}_n\hat {\vec S}_{n+1} +\delta
{\hat S}_n^z{\hat S}^z_n)
\end{equation}

(single-ion anisotropy). \\
Here $\hat S_n^x, \hat S_n^y, \hat S_n^z$ are the spin operators acting
at a site n, $\delta$ is the anisotropy coefficient.

In the one-dimensional case with spin $s=1/2$, the study can be done
completely [1]. Higher spin models (and more dimensional) require
approximate treating.  One of them we use later is the so-called trial
function method. Based on minimization of the Hamiltonian with respect to
a set of trial functions this method is in fact very sensitive to the
choice of these functions. There are certain thoughts (ideas) and even
theorems aiding in the search for them.  Usually these ideas are based on
symmetry properties of the system under consideration (e.g., the
Coleman-Palais theorem). Symmetry turns out to be the most powerful and
effective tool here. This is why we choose generalized coherent states
(for definition see later and e.g. [2]) as the trial functions for models
(1) and (2).

The paper is organized as follows. In the second part, we construct GCS
defined on the homogeneous (complex projective) spaces
$SU(2j+1)/SU(2j)\otimes U(1) \equiv {\bf CP}^{2j}$. The emphasis will be
on the $SU(2)/U(1)$ GCS, reducing the quantum description to classical
Landau-Lifshitz phenomenology. In the other cases, we shall only be
interested in GCS in the faithful (fundamental) representation of
corresponding groups. Also GCS constructed on the noncompact group
$SU(1,1)$ will be presented in order to treat the antiferromagnet models
($+$ sign in (1) and (2)).

In the third part, classical counterparts of the spin operators and their
products are obtained and discussed.

The fourth part is devoted to deriving classical lattice Hamiltonians for
various spin values, {\it s}, starting from (1) and (2). Here, we note that
the dimensions of the spin phase space at a lattice site coincides with that
of the corresponding coset space. Indeed, we have for an arbitrary spin state
at a site,
\begin{equation}
\mid\Psi> =\sum_{m=1}^{2s+1}c_m\mid\psi_m>
\end{equation}

where $\mid\psi_m>$ are pure spin states, e.g., $\mid\psi_m>=\mid s,m>$, and
$c_m$ are complex constants. The vector $\mid\Psi>$ is defined up to an
arbitrary phase:$\mid\Psi> \simeq \mid\Psi>e^{i\theta}$ and should satisfy
the normalization condition:
$$<\Psi\mid\Psi>=1.$$ \\
These two real conditions reduce the dimension of the spin phase space,
{\bf S}, by two:
\begin{equation}
dim{\bf S}=2(2s+1)-2=4s .
\end{equation}

The dimension of ${\bf CP}^{2s}$ is
\begin{equation}
dim{\bf CP}^{2s}= (2s+1)^2-4s^2-1=4s ,
\end{equation}

so
$$ dim{\bf CP}^{2s}= dim{\bf S}$$

and ${\bf CP}^{2s}$ GCS provide a complete description of the corresponding
model.

The fifth part deals with classical lattice equations of motion derived by
considering the quantum probability amplitude.

The sixth part considers  continuum models resulting from the classical
lattice models. Some properties of these continuum models are discussed.

\item {\large {GCS defined on the coset space $SU(2j+1)/SU(2j)\otimes U(1)$ }}

Generalized coherent states (GCS), the extension of the well-known Glauber
coherent states related to the Heisenberg-Weyl group, are defined and
discussed in the book [3]. Here we chose them as the trial functions for,
at least, three reasons:

1) Our aim is semiclassical description of spin quantum models and GCS usually
give such since they minimize the uncertainty relation.

2) They provide a minimal description (no extra parameters) as we'll see later
on.

3) They are taken according to the Hamiltonian symmetry demands.

Therefore our concern in what follows will be with GCS defined on the
complex projective spaces, ${\bf CP}^{2j}$. The simplest is the sphere, ${\bf
S}^2= {\bf CP}^1= SU(2)/U(1)$. To its points are related so called spin
coherent states discussed in detail in many papers and books (see, e.g. [2]and
[3]).  These states can be parametrized by the points of a real sphere, ${\bf
S}^2$ or, via the stereographic projection onto the complex plane, $\bf C$, by
its points:  \begin{equation} \mid\Psi>=e^{\alpha {\hat S}^+-{\bar\alpha}{\hat
S}^-}|0>=(1+|\psi|^2)^{-j}e^ {\psi{\hat S}^+}|0> \end{equation} with $${\hat
S}^{\pm}={\hat S}^x \pm i{\hat S}^y  \quad and \quad \psi=\frac{\alpha}
{|\alpha|}tan{|\alpha|},$$
where $\alpha$ and $\psi$ are complex numbers; $|0>=|j,-j>$, the ground state
and j defines the unitary representation of the group, $SU(2)$, and hence the
spin value: $s=j$. By stereographic projection, we have $\psi
=-tan{\frac{\theta}{2}} e^{i\phi}, |\alpha|\leq \frac{\pi}{2}$. Thus, the set
of GCS (6) has spherical symmetry and is (in principle) valid for any spin
value.  It means that system (6) can be used as the trial function basis when
the symmetry of the quantum Hamiltonian is very close to spherical symmetry:
$\delta =0$ or $\delta \ll 1$.

GCS for other groups are constructed using their fundamental representation:
\begin{equation}
\mid\Psi>=exp\bigl\{\sum_i^{2s}(\zeta_i{\hat T}_i^+ -{\bar\zeta}_i{\hat T}_i^-)
\bigr\}|0>=(1+\sum_i^{2s}|\psi_i|^2)^{-\frac{1}{2}}\bigl\{|0>+\sum_i^{2s}
\psi_i|i>\bigr\}
\end{equation}
where ${\hat T}_i^+$ and ${\hat T}_i^-$ are generators of the SU(2j+1) group
in the fundamental representation and
$$\psi_i=\frac{\zeta_i}{|\zeta|}tan{|\zeta|}, \qquad |\zeta|=\sqrt{\sum_i^{2s}
|\zeta_i|^2}$$
$$|0>=(0,...,0,1)^{tr},\quad  |i>=(0,...,0,1,\underbrace{0,...,0}_i)^{tr}$$

For example, in the ${\bf CP}^2$ case we have $G=SU(3)$, the coset space
$G/H=SU(3)/SU(2)\otimes U(1) $ and
\begin{equation}
\mid\Psi>=(1+|\psi_1|^2+|\psi_2|^2)^{-\frac{1}{2}}\{|0>+\psi_1|1>+\psi_2|2>\}
\end{equation}
Let us consider GCS for $s=1$ systems. In this case, states (6) are
\begin{equation}
\mid\Psi>=\frac{1}{1+|\psi|^2}\{|0>+\sqrt{2}\psi|1>+\psi^2|2>\}.
\end{equation}
These states are parametrized by a complex function, $\psi$, i.e. this system
lives on a two-dimensional real manifold (the spin phase space). States (8)
are parametrized by two complex functions, $\psi_1$ and $\psi_2$, so the
system lives on a four-dimensional real manifold (the spin phase space). We
saw earlier that the spin phase space in the $s=1$ case was four-dimensional
and the second system in this regard was more appropriate.

One can easily check that under the conditions
$$\psi_1=\sqrt{2}\psi, \quad \psi_2=\psi_1^2/2 $$
both systems coincide. It means that the first 2-d manifold is just a section
of the second 4-d manifold, under the constraint
\begin{equation}
\psi_2=\frac{1}{2}\psi_1^2
\end{equation}

An analogous projection will occur for $s=\frac{3}{2}$ where
\begin{equation}
\mid\Psi>=(1+|\psi_1|^2+|\psi_2|^2+|\psi_3|^2)^{-\frac{1}{2}}\bigl(|0>+
\psi_1|1>+\psi_2|2>+\psi_3|3>\bigr)
\end{equation}
and the constraint determining the two-dimensional SU(2) section is
\begin{equation}
\psi_2=\frac{1}{\sqrt{3}}\psi_1^2, \quad \psi_3=\frac{1}{3\sqrt{3}}\psi_1^3
\end{equation}

Here it is of use to stress that, in the case $s=1$, the states (8) can be
rewritten in the so-called angle parametrization (see, for example [4]):
\begin{equation}
|\Psi>=e^{-i\phi{\hat S}^z}e^{-i\theta{\hat S}^y}e^{i\gamma{\hat S}^z}
e^{2ig{\hat Q}^{xy}}|0>=c_0|0>+c_1|1>+c_2|2>
\end{equation}
where
\begin{displaymath}
{\hat Q}^{xy}=\frac{i}{2}
\left( \begin{array}{ccc}
0 & 0 & -1 \\
0 & 0 & 0 \\
-1 & 0 & 0
\end{array} \right)
\end{displaymath}
\begin{eqnarray}
c_o & = & e^{i\phi}(e^{-i\gamma}\sin^2{\frac{\theta}{2}}\cos g+e^{i\gamma}
\cos^2{\frac{\theta}{2}}\sin g) \nonumber\\
c_1 & = & \frac{\sin\theta}{\sqrt{2}}(e^{-i\gamma}\cos g+e^{i\gamma}\sin g)
\nonumber\\
c_2 & = & e^{-i\phi}(e^{-i\gamma}\cos^2{\frac{\theta}{2}}\cos g +e^{i\gamma}
\sin^2{\frac{\theta}{2}}\sin g)
\end{eqnarray}

Two angles, $\theta$ and $\phi$, define the orientation of the classic spin
vector. The angle, $\gamma$, is the rotation of the quadrupole moment about
the spin vector. The parameter, g, defines change of the spin vector magnitude
and that of the quadrupole moment.

Finally, we give GCS defined on the noncompact manifold $SU(1,1)/U(1)$ the two
sheet hyperboloid ${\bf S}^{1,1}$:
\begin{equation}
|\zeta>=(1-|\zeta|^2)^ke^{\zeta{\hat K}^+}|0>
\end{equation}
where again $\zeta$ is a complex number, ${\hat K}^+={\hat K}^x +i{\hat K}^y$,
and $|0>$ is the ground state: ${\hat K}^-|0> =0$, i.e. $|0>=|k,k>$.

\item {\large {Averaged spin operators and their products}}

Here we consider classical counterparts of the spin operators and their
products contained in the Hamiltonians (1) and (2).

The vector
\begin{equation}
\vec S=<\Psi|\hat {\vec S}|\Psi>
\end{equation}
can be regarded as a classical spin vector, and
\begin{equation}
Q^{ij}=<\Psi|\hat S^i \hat S^j|\Psi>
\end{equation}
as a component of the quadrupole moment.

Because the spin operators at different lattice sites commute, we have for
all such products
\begin{equation}
<\Psi|\hat S_n^i \hat S_{n+1}^k|\Psi>=<\Psi|\hat S_n^i|\Psi><\Psi|\hat
S^k_{n+1}|\Psi>,
\end{equation}
where
$$|\Psi>=|\Psi>_n|\Psi>_{n+1}.$$

Corresponding expressions for the ${\bf CP}^{1}$ GCS are well-known [2,3]:
\begin{equation}
S^+=\bar S^-=2j\frac {\bar \psi}{1+|\psi|^2}, \qquad
S^z= -j\frac{1-|\psi|^2}{1+|\psi|^2}
\end{equation}
\begin{eqnarray}
Q^{zz} & = & <\Psi|\hat S^z\hat S^z|\Psi>=\frac{j^2(1-|\psi|^2)+2j|\psi|^2}
{(1+|\psi|^2)^2}= \nonumber\\
& = & (S^z)^2\bigl[1+\frac{2}{j}\frac{|\psi|^2}{(1-|\psi|^2)^2}\bigr]
\equiv(S^z)^2+\frac{1}{2j}S^+S^-.
\end{eqnarray}
The quasiclassical limit requires
$$2j\gg \frac{S^+S^-}{(S^z)^2}.$$
For the same reason for
\begin{equation}
Q^{+-}=S^+S^-\bigl(1+\frac{1}{2j}\frac{j+S^z}{j-S^z}\bigr) \quad or
\quad Q^{-+}=S^+S^-\bigl(1+\frac{1}{2j}\frac{j-S^z}{j+S^z}\bigr)
\end{equation}
the quasiclassical limits mean
$$2j\gg\frac{j+S^z}{j-S^z} \quad or \quad 2j\gg\frac{j-S^z}{j+S^z}.$$

One can also see that any quadrupole moment component, $Q^{ij}$, is expressed,
though nonlinearly, through two components of the spin vector $\vec S$. Higher
order moments, $Q^{ijk}$, and so on are readily shown to be expressed, in a
similar way.

Components of the classical spin vector, $\vec S$, and of the quadrupole
moment, $Q^{ij}$, for other sets of GCS are less known, though straightforward:
\begin{equation}
S^+=\bar S^-=\sqrt 2\frac{\bar\psi_1+\psi_1\bar\psi^2}{1+|\psi_1|^2+|\psi_2|^2} \qquad (j=1)
\end{equation}
\begin{equation}
S^z=\frac{|\psi_2|^2-1}{1+|\psi_1|^2+|\psi_2|^2} \qquad (j=1)
\end{equation}
\begin{equation}
Q^{zz}=\frac{1+|\psi_2|^2}{1+|\psi_1|^2+|\psi_2|^2} \qquad (j=1)
\end{equation}
(all other components of $Q^{ij}$ can be expressed in terms of $S^+,S^z$ and
$Q^{zz}$).
\begin{equation}
S^+=\bar S^-=\frac{\sqrt 3\psi_2\bar\psi_3+2\psi_1\bar\psi_2+\sqrt 3\bar\psi_1}
{1+|\psi_1|^2+|\psi_2|^2+|\psi_3|^2}  \qquad (j=\frac{3}{2})
\end{equation}
\begin{equation}
S^z=\frac{3|\psi_3|^2+|\psi_2|^2-|\psi_1|^2-3}{1+|\psi_1|^2+|\psi_2|^2+
|\psi_3|^2}  \qquad (j=\frac{3}{2})
\end{equation}
\begin{equation}
Q^{zz}=\frac{1}{4}\frac{9+|\psi_1|^2+|\psi_2|^2+9|\psi_3|^2}{1+|\psi_1|^2+
|\psi_2|^2+|\psi_3|^2}  \qquad (j=\frac{3}{2})
\end{equation}

We note here that the averaged Casimir operators are
$$\hat{\bf C_2}=\frac{1}{2}(Q^{+-}+Q^{-+})+Q^{zz}=j(j+1) \quad \mbox{ for the
${\bf CP}^1$ GCS, }  $$
$$\hat{\bf C_2}=j(j+1)=2 \quad \mbox{ for the ${\bf CP}^2$ GCS }  $$
 and
$$\hat{\bf C_2}=j(j+1)=\frac{15}{4} \quad \mbox{ for the ${\bf CP}^3$ GCS. }$$

More important to notice that for the ${\bf CP}^1$ GCS,
\begin{equation}
\vec S^2=const=j^2.
\end{equation}
For the ${\bf CP}^2$ GCS,
\begin{equation}
\vec S^2+q^2=1 \qquad (j=1)
\end{equation}
with
\begin{equation}
q^2=Q^{z+}Q^{z-}+Q^{+z}Q^{-z}+Q^{++}Q^{--}+(1-Q^{zz})^2
\end{equation}
The analogous formula for the ${\bf CP}^3$ GCS is
$$\vec S^2+q^2+t^2=const,$$
with more cumbersome expressions for $q^2$ involving $Q^{ij}$ and $t^2$
through $Q^{ijk}$.

We emphasize that, in the ${\bf CP}^1$ case (j=1) by use of the angle
parametrization (14), one has [5]:
\begin{eqnarray}
S^+ & = & e^{i\phi}\cos{2g}sin\theta ,  \nonumber \\
S^z & = & \cos{2g}cos\theta   \nonumber, \\
q^2 & = & \sin^2{2g}   \nonumber, \ and \\
Q^{zz} & = & 1-\frac{1}{2}sin^2\theta+\frac{1}{2}sin^2\theta\cos{2\gamma}
\sin{2g}.
\end{eqnarray}
The identity $S^2+q^2=1$ is trivially satisfied.

Finally, we give the expressions for the operators $\hat K^{\pm}$ and
$\hat K^z$ (the generators of the group SU(1,1)) averaged over the
corresponding ${\bf L}^{1,1}=SU(1,1)/U(1)$  GCS:
\begin{equation}
K^+=\bar K^-=<|\hat K^+|>=2k\frac{\bar \zeta}{1-|\zeta|^2}
\end{equation}
\begin{equation}
K^z=k\frac{1+|\zeta|^2}{1-|\zeta|^2}
\end{equation}
\begin{eqnarray}
K^{zz}=<|\hat K^z\hat K^z|>=(K^z )^2\bigl(1+\frac{2}{k}\frac{|\zeta|^2}{(1+
|\zeta|^2)^2}\bigr)=(K^z)^2+\frac{1}{2k}K^+K^- \\
K^{+-}=K^+K^-(1+\frac{1}{2k}|\zeta|^2), \qquad K^{-+}=K^+K^-(1+\frac{1}{2k}
|\zeta|^{-2})
\end{eqnarray}
such that the averaged Casimir operator is
$${\bf C}_2=\frac{1}{2}(K^{+-}+K^{-+})-K^{zz}=k(1-k)$$
and the classical pseudovector $\vec K=(K^x,K^y,K^z)$ obeys the condition:
\begin{equation}
K^+K^--(K^z)^2=-k^2
\end{equation}
and lies on the two sheet hyperboloid ${\bf S}^{1,1}$ (we omit the details
which appear in [2]).

\item {\large {Classical lattice Hamiltonians}}

In this part, we derive classical lattice Hamiltonians which are just
Hamiltonians (1) and (2) averaged over different GCS.\\
1) First we use the spin coherent states and consider model (1). As was
already mentioned the spin operators at neighboring sites commute, so the
coherent state of the whole lattice is
\begin{equation}
|\Psi>=\prod_n|\Psi>_n
\end{equation}

Averaging (1) with (37) and using eqs. (19) in the ferromagnetic case one
has
\begin{eqnarray}
H_e & = & <\Psi|\hat H_e|\Psi>=-J\sum_n(\vec S_n\vec S_{n+1}+\delta S_n^z
S_{n+1}^z) \nonumber \\
& = & -j^2J\sum_n\frac{2(\bar\psi_n\psi_{n+1}+c.c.)+(1+\delta)(1-|\psi_n|^2)
(1-|\psi_{n+1}|^2)}{(1+|\psi_n|^2)(1+|\psi_{n+1}|^2)}
\end{eqnarray}
the classical lattice Hamiltonian which in the continuum limit (d=1) becomes
\begin{equation}
H_e=-j^2JN+J\int
\limits_{-\infty}^{\infty}\frac{dx}{a_o}\left(\frac{a_o^2}{2}\vec S_x\vec
S_x-\delta S^zS^z\right) \end{equation} for the $\sigma-model$ representation,
or \begin{equation} H_e=-const+2j^2Ja_o\int
\limits_{-\infty}^{\infty}dx\frac{|\psi_x|^2+\rho |\psi|^2}{(1+|\psi|^2)^2}
\end{equation}
for the stereographic projection with $\delta=\frac{a_o^2}{2}\rho$. $N$ is
the total number of lattice sites.

From eq.(40), we see that classical energy above the ground state (a large
negative constant) is positive and classical excitations will possess positive
energy. On the contrary, in the antiferromagnetic case direct application of
the spin GCS leads to excitations with negative energy, i.e. such a system
should be unstable (more correct, the quantum vacuum over which we construct
the excitations is unstable). Apparently, this is the reason why the search
for the vacuum in the antiferromagnetic case is such a complicated problem.
To avoid this difficulty and provide excitations with positive energy,
following [2], we use the following trick. Rewrite (1) via operators of the
$su(1,1)$ algebra
\begin{equation}
\hat K^{\pm}=i\hat S^{\pm}, \qquad \hat K^z=\hat S^z.
\end{equation}
Then we have the pseudospin representation for antiferromagnet:
\begin{equation}
\hat H_e=-J\sum_n\left[\frac{1}{2}(\hat K^+_n\hat K^-_{n+1}+h.c.)-\hat K^z_n\hat
K^z_{n+1}(1+\delta)\right].
\end{equation}

Now, we treat this model applying the above scheme and using ${\bf L}^{1,1}$
GCS to obtain the classical lattice model:
\begin{equation}
H_e=-k^2J\sum_n\frac{2(\bar\zeta_n\zeta_{n+1}+\zeta_n\bar\zeta_{n+1})-
(1+\delta)(1+|\zeta_n|^2)(1+|\zeta_{n+1}|^2)}{(1-|\zeta_n|^2)(1-
|\zeta_{n+1}|^2)}.
\end{equation}
The continuum limits are:
\begin{eqnarray}
H_e & = & -J\sum_n\vec K^2+\frac{a_o}{2}J\int dx(\vec K_x\vec K_x+\rho K^zK^z)
\nonumber\\
& = & Jk^2N+\frac{a_o}{2}\int dx(\vec K_x\vec K_x+\rho K^zK^z)
\end{eqnarray}
for the $\sigma-model$ representation, or
\begin{equation}
H_e=const+2k^2a_oJ\int dx\frac{|\zeta_x|^2+\rho|\zeta|^2}{(1-|\zeta|^2)^2}
\end{equation}
for the stereographic projection.

Thus, we avoid the problem of excitations with negative energy but obtain,
instead, the problem of treating noncompact groups and manifolds
($\sigma-model$ representation) or singular expressions (stereographic
projection).

Models (38)-(40) and (43)-(45) can be regarded as appropriate if the
anisotropy constant $\delta$ is very small: $\delta\ll 1$. Then, symmetries
of the quantum Hamiltonians and of GCS manifolds should be very close:
spherical for the ferromagnet (1) and pseudospherical for antiferromagnet (1).
While taking to the continuum limit, we require, $a_o/\lambda\ll 1 $ where
$\lambda$ is the wavelength considered.

The same procedure can be applied to models (2) to give
\begin{eqnarray}
H_s & = & <\Psi|\hat H|\Psi>=-J\sum_n(\vec S_n\vec S_{n+1}+\delta Q_n^{zz})
\nonumber\\
& = & -J\sum_n\left\{\vec S_n\vec S_{n+1}+\delta
\left[(S_n^z)^2+\frac{1}{2j}S_n^+ S_n^-\right]\right\},
\end{eqnarray}
or
\begin{eqnarray}
H_s = & - &   j^2 J\sum_n[\frac{
2(\bar\psi_n\psi_{n+1}+c.c.)+(1-|\psi_n|^2) (1-|\psi_{n+1}|^2) }{
(1+|\psi_n|^2)(1+|\psi_{n+1}|^2) }+ \nonumber\\
& + &  \delta\frac{(1-|\psi_n|^2)^2+\frac{2}{j}|\psi_n|^2}{ (1+|\psi_n|^2)^2}].  \end{eqnarray}
In the continuum limit, \begin{equation}
H_s=-j^2JN\left(1+\frac{\delta}{2j}\right)+\frac{a_o}{2}J\int dx\left[\vec
S_x\vec S_x- \delta(1-\frac{1}{2j})(S^z)^2\right] \end{equation} for the
$\sigma-model$ representation, or
\begin{equation}
H_s=-j^2JN\left(1+\frac{\delta}{2j}\right)+2a_oj^2J\int
dx\frac{|\psi_x|^2+\rho_1|\psi|^2} {(1+|\psi|^2)^2}, \quad
\rho_1=\rho\left(1-\frac{1}{2j}\right) \end{equation} for the stereographic
projection.

These formulas imply both systems being equivalent, up to renormalization of
the following constants: the ground state energy level and the anisotropy rate. Hence, the classical dynamics of both systems is the same (with the exception
of the $j=\frac{1}{2}$ case).

If we are given with Hamiltonian (2) sign (-), then
\begin{equation}
H_s=k^2JN(1-\frac{\delta}{2k})+\frac{a_o}{2}J\int dx[\vec K_x\vec K_x+\delta(1+
\frac{1}{2k})(K_z)^2]
\end{equation}
for the $\sigma-model$ representation, or
\begin{equation}
H_s=k^2JN(1-\frac{\delta}{2k})++2a_ok^2J\int dx\frac{|\zeta_x|^2+\rho_2
|\zeta|^2}{91-|\zeta|^2)^2}, \quad \rho_2=\rho(1+\frac{1}{2k})
\end{equation}
for the stereographic projection.

Again both systems are equivalent.

Let us now proceed to other GCS. \\
2) Here we use ${\bf CP}^2$ GCS for treating model (1) with $s=1$. Then we
have the same expressions for the Hamiltonians in the $\sigma-model$
representation,viz. (38),(39) and (44). In the complex representation, we
have the lattice Hamiltonian:
\begin{eqnarray}
H_e= & - & j^2J\sum_n\bigl[\frac{(\bar\psi_{1n}+\psi_{1n}\bar\psi_{2n})
(\psi_{1n+1}+\bar\psi_{1n+1}\psi_{2n+1})+c.c.}{(1+|\psi_{1n}|^2+|\psi_{2n}|^2)
(1+|\psi_{1n+1}|^2+|\psi_{2n+1}|^2)}+  \nonumber\\  & + &  \frac{(1+\delta)
(1-|\psi_{2n}|^2)(1-|\psi_{2n+1}|^2)}{(1+|\psi_{1n}|^2+|\psi_{2n}|^2)(1+
|\psi_{1n+1}|^2+|\psi_{2n+1}|^2)}\bigr].
\end{eqnarray}
The corresponding continuum expression is very complicated and we do not
give it here for the general case. But to understand the difference between
the two models, we must study their {\it ground states}. It is natural to
suggest that, at least in the ferromagnetic case, this state will be close
to the one for which
\begin{equation}
\psi_n=\psi_l, \quad at \quad n\not= l.
\end{equation}
Then, Hamiltonian of model (38) is as follows
\begin{equation}
H_e=-J\sum_n(\vec S_n^2+\delta(S_n^z)^2)=-j^2JN-\delta J\sum_n(S_n^z)^2
\end{equation}
and the energy assumes its minimum when \\
a) $\delta>0$ (easy-axis model)
\begin{equation}
\vec S=(0,0,S^z)\equiv (0,0,j) :\quad E=-j^2JN(1+\delta).
\end{equation}
b) $\delta<0$ (easy-plane model)
\begin{equation}
\vec S=(S^x,S^y,0) ;\qquad E=-j^2JN.
\end{equation}

For the ${\bf CP}^2$ model, we have (due to $s^2+q^2=1$ at j=1)
\begin{equation}
H_e=-J\sum_m(\vec S_n^2+\delta(S_n^z)^2)=-JN-J\sum_n(\delta(S_n^z)^2-q_n^2)
\end{equation}
with $q_n^2$ given by (30).

The ground states again are the states with $q_n^2=0$. However, in looking
for excitations in the frame of the ${\bf CP}^1$ model, the term,
\begin{equation}
H_{1e}=-J\sum_n\delta(S_n^z)^2,
\end{equation}
is varied. In the ${\bf CP}^2$ case,
\begin{equation}
H_{2e}=-J\sum_n[\delta(S_n^z)^2-q_n^2]
\end{equation}
i.e. here an additional term, $\sum_nq_n^2$, proportional to quadrupole
moment and of effective anisotropy in nature, appears. Moreover in the first
case classical dynamics is orientation dynamics such that the classical spin
vector being of constant value just alters its direction (lies on the sphere
${\bf S}^2$). In the second case the vector can alter its value (along with
quadrupole moment) as well as the direction.

The ${\bf CP}^3$ model, besides the quadrupole moment, q, has an octupole
moment, $t$, and, hence, one more term:
\begin{equation}
H_{3e}=-J\sum_n[\delta(S_n^z)^2-q_n^2-t_n^2]
\end{equation}
and so on.

Consider model (2). Here in the ${\bf CP}^1$ case, we have Hamiltonian (58)
with $\delta_1=\delta(1-\frac{1}{2j})$. But for the ${\bf CP}^2$ model,
\begin{equation}
H_s=-J\sum_n(\vec S_n\vec S_{n+1}+\delta Q_n^{zz})
\end{equation}
and, in the vicinity of the ground state, we have
\begin{equation}
H_s=-JN-J\sum_n(\delta Q_n^{zz}-q_n^2),
\end{equation}
the Hamiltonian containing a pair product of spin operators.

Let us consider again the ground state of the easy-axis type. We have
(see (22))
$$S^+=\sqrt 2\frac{\bar\psi_1+\psi_1\bar\psi_2}{1+|\psi_1|^2+|\psi_2|^2}=0$$
if \\
1. $\psi_1=\psi_2=0$ and $S^z=1, q^2=0$; this state is similar to the
${\bf CP}^1$ case.

2. $\psi_1=0$, $\psi_2$ is arbitrary:
$$S^z=\frac{|\psi_2|^2-1}{|\psi_2|^2+1}<1,\quad q^2=4\frac{|\psi_2|^2}
{1+|\psi_2|^2}.$$

3. by expressing $\psi_n=|\psi_n|e^{i\phi_n}$ one has $e^{-i\phi_1}+
|\psi_2|e^{i(\phi_1-\phi_2)}=0$ or

$\psi_1$ is arbitrary and $\psi_2=e^{i(2\phi_1-\pi)}$.

It is easy to check that the first solution has the minimal energy:
$$E=-JN(1+\delta), \quad \delta>0.$$
The same result can be expressed in terms of the angle parametrization (14),
where
\begin{equation}
H_s=-\sum_n[\cos^2{2g}+\delta(1-\frac{1}{2}sin^2{\theta}(1-\cos{2\gamma}
\sin{2g}))]
\end{equation}
and if $\delta>0$ we have $sin\theta=0$ or $\theta=0$ and $g=0$, i.e. the
known ${\bf CP}^1$ easy-axis vacuum.

Longer calculations show that for $\delta<0$, we arrive at the easy-plane
ground state: $S^z=0$ but now there is a spin value reduction:
\begin{equation}
S^2=1-\frac{1}{16}\delta^2, \qquad |\delta|< 4
\end{equation}
and we have the minimal energy
\begin{equation}
 E=-JN(1-\frac{1}{2}\delta+\frac{1}{16}\delta^2)
\end{equation}
at
\begin{equation}
sin^2g=\frac{1}{16}\delta^2 ,\qquad \theta=0.
\end{equation}
The same expressions in the complex parametrization are readily given by
\begin{eqnarray}
|\psi_2|^2 & = & 1 \\
|\psi_1|^2 & = & 2\frac{4-\delta}{4+\delta} \\
\phi_2 & = & 2\phi_1
\end{eqnarray}

For the ${\bf CP}^3$ case, it is easy to show that ground states for model
(1) are the same in terms of the classical spin vectors and $q^2=g=0$.
Moreover, here again $\partial_tS^2=0$. In this case model (2) requires
further study.

\item{\large {Classical lattice equations of motion}}

To obtain classical equations of motion we, following [2], consider the
transition amplitude from the CS $|\psi>$ at time $t$ to the state
$|\psi_1>$ at instant $t_1$:  $$
P(\psi_1,t_1|\psi,t)=<\psi_1|exp\{-\frac{i}{\hbar}\hat H(t_1-t)\}|\psi>$$
Dividing up the interval $t_1-t$ into $n$
equal subintervals $\epsilon=\frac{1}{n}(t_1-t)$ and passing to the limit $n\rightarrow\infty$, we have
$$P(\psi_1,t_1|\psi,t)= \lim_{n \to \infty}<\psi_1|(1-\frac{i}{\hbar}\hat
H\epsilon_1)(1-\frac{i}{\hbar}\hat H\epsilon_2)\ldots(1-\frac{i}{\hbar}\hat
H\epsilon_n)|\psi_n>$$
By using the fact that the GCS obey the relation
$$\int d\mu_i(\psi)|\psi><\psi|=I,$$
we then obtain

\begin{eqnarray}
P & = & \lim_{n \to \infty}\int\ldots\int [\prod_{k=1}^{n-1}d\mu(\psi_k)]
\prod_{k=1}^n<\psi_k|(1-\frac{i}{\hbar}\hat H\epsilon)|\psi_{k-1}> \nonumber\\
& = & \lim_{n \to \infty}\int\ldots\int[\prod_{k=1}^{n-1}d\mu(\psi_k)]
\prod_{k=1}^n<\psi_k|\psi_{k-1}>(1-\frac{i\epsilon}{\hbar}\frac{<\psi_k|\hat H|
\psi_{k-1}>}{<\psi_k|\psi_{k-1}>}),    \nonumber
\end{eqnarray}
with $\psi_o=\psi$ and $\psi_n=\psi_1$. In the limit $\epsilon\rightarrow 0$
we have
$$1-\frac{i\epsilon}{\hbar}\frac{<\psi_k|\hat H|\psi_{k-1}>}
{<\psi_k|\psi_{k-1}>}=exp\{-\frac{i\epsilon}{\hbar}\frac{<\psi_k|\hat
 H|\psi_{k-1}>}{<\psi_k|\psi_{k-1}>}\}$$
Now since $\epsilon\ll 1$, we have $\psi_{k-1}=\psi_k-\Delta \psi_k$ and then
\begin{eqnarray}
<\psi_k|\psi_{k-1}> & = & 1-\partial_{\psi'}<\psi_k|\psi'>|_{\psi'=\psi_k}
\Delta\psi_k-\partial_{\bar\psi'}<\psi_k|\psi'>|_{\psi'=\psi_k}
\Delta\bar\psi_k \nonumber \\
& = & 1+\frac{1}{2}\frac{\psi_{1k}\Delta\bar\psi_{1k}+\psi_{2k}\Delta\bar
\psi_{2k}-c.c.}{1+|\psi_1|^2+|\psi_2|^2} +O[(\Delta\psi_i)^2], \nonumber
\end{eqnarray}
or
$$ <\psi_k|\psi_{k-1}>\cong exp\{\frac{1}{2}\frac{\psi_{1k}\Delta\bar\psi_{1k}+
\psi_{2k}\Delta\bar\psi_{2k}-c.c.}{1+|\psi_1|^2+|\psi_2|^2}\}, $$
then
\begin{eqnarray}
\prod_{k=1}^n<\psi_k|\psi_{k-1}>=exp\{\frac{1}{2}\sum\frac{\psi_{1k}\Delta
\bar\psi_{1k}+\psi_{2k}\Delta\bar\psi_{2k}-c.c.}{1+|\psi_1|^2+|\psi_2|^2}\}
\nonumber \\
=exp\{\frac{1}{2}\int dt\frac{\psi_{1k}\frac{d}{dt} \bar\psi_{1k}+\psi_{2k}
\frac{d}{dt} \bar\psi_{2k}-c.c.}{1+|\psi_1|^2+|\psi_2|^2}\}. \nonumber
\end{eqnarray}

Finally, combining all the expressions, we obtain
$$P(\psi_1,t_1|\psi,t)=\int {\it D}\mu (\psi)exp\left\{\frac{i}{\hbar}\int
\limits_{t}^{t'}d\tau{\cal L}(\psi_1,\psi_2,\bar\psi_1,\bar\psi_2)\right\},$$
where
\begin{equation}
{\cal L}=\frac{\frac{i\hbar}{2}}{1+|\psi_1|^2+|\psi_2|^2}\sum_{l=1}^{2}
(\psi_l\frac{d}{dt}\bar\psi_l-c.c.)-{\cal H}
\end{equation}
with
\begin{equation}
{\cal H}= <\psi_k|\hat H|\psi_k>
\end{equation}
at a lattice site.

The total classical lattice Lagrangian is the sum over all sites
\begin{equation}
L=\sum_{n=1}^N L_n
\end{equation}
where $ L_n={\cal L}$ at site $n$.
By varying (72) with respect to $\bar\psi_1$ and $\bar\psi_2$,
one has at a site
\begin{eqnarray}
\frac{i}{1+|\psi_1|^2+|\psi_2|^2}\{\dot\psi_1(1+|\psi_2|^2)-\psi_1\bar
\psi_2\dot\psi_2\}=\frac{\partial {\cal H}}{\partial\bar\psi_1} \nonumber \\
\frac{i}{1+|\psi_1|^2+|\psi_2|^2}\{\dot\psi_2(1+|\psi_1|^2)-\psi_2\bar
\psi_1\dot\psi_1\}=\frac{\partial {\cal H}}{\partial\bar\psi_2} \nonumber
\end{eqnarray}
or for the lattice
\begin{equation}
i\dot\psi_{1n}=\left(1+|\psi_{1n}|^2+|\psi_{2n}|^2\right)\left((1+|\psi_{1n}|^2)\frac{\partial H}{\partial\bar\psi_{1n}}+\psi_{1n}\bar\psi_{2n}\frac{\partial H} {\partial
\bar\psi_{2n}}\right), \quad (1\rightleftharpoons 2),
\end{equation}
where $ H=\sum_n H_n$ is the classical lattice Hamiltonian.

In the simplest case of weak exchange anisotropy, we have (${\bf CP}^1$ case)
\begin{equation}
\frac{1}{j^2J} H_{em} =-2\frac{\bar\psi_n\psi_{n+1}+c.c.}{(1+|\psi_n|^2)(1+
|\psi_{n+1}|^2)}-(1+\delta)\frac{(1-|\psi_n|^2)(1-|\psi_{n+1}|^2)}{(1+
|\psi_n|^2)(1+|\psi_{n+1}|^2)}
\end{equation}
and the equations of motion (see [2])
$$i\dot\psi_n=(1+|\psi_n|^2)^2\frac{\partial H}{\partial\bar\psi_n}$$
i.e.
\begin{eqnarray}
\frac{i}{2}\dot\psi_n & = & \frac{\psi_{n-1}}{1+|\psi_{n-1}|^2}+\frac{
\psi_{n+1}}{1+|\psi_{n+1}|^2} \nonumber \\
 & - & \psi_n^2\left(\frac{\bar\psi_{n-1}}{1+|\psi_{n-1}|^2}+\frac{\bar
\psi_{n+1}}{1+|\psi_{n+1}|^2}\right) \nonumber \\
 & - & 2(1+\delta)\psi_n\frac{1-|\psi_{n-1}|^2|\psi_{n+1}|^2}{(1+
|\psi_{n-1}|^2)(1+|\psi_{n+1}|^2)},
\end{eqnarray}
which, in small amplitude region, is
\begin{eqnarray}
\frac{i}{2}\dot\psi_n & = & \psi_{n-1}+\psi_{n+1}-2(1+\delta)\psi_n \nonumber\\
 & - & (\psi_{n-1}|\psi_{n-1}|^2+\psi_{n+1}|\psi_{n+1}|^2) \nonumber \\
 & - &  \psi_n^2(\bar\psi_{n-1}+\bar\psi_{n+1})+
\psi_n(|\psi_{n-1}|^2+\psi_{n+1}|^2).
\end{eqnarray}
the equation of the so-called $\phi_4$-theory.
It is easy to verify that eq.(76) possesses two ground states solutions:
$$\psi_n=\psi_{n\pm 1}$$
then
$$-2\delta\psi_n\frac{1-|\psi_n|^4}{(1+|\psi_n|^2)^2}=0$$
and
\begin{eqnarray}
\psi_n & = & 0 \qquad \mbox{being the  easy-axis vacuum state} \\
|\psi_n|^2 & = & 1 \qquad \mbox{being the easy-plane vacuum state.}
\end{eqnarray}
Eq. (71) only has the easy-axis vacuum.

For $spin=\frac{3}{2}$, one has instead of (73)
\begin{eqnarray}
i\dot\psi_{1n}=(1+|\psi_{1n}|^2+|\psi_{2n}|^2+|\psi_{3n}|^2)\{(1+
|\psi_{1n}|^2)\frac{\partial H}{\partial\bar\psi_{1n}}+
                                             \nonumber\\
  +\psi_{1n}\bar\psi_{2n}\frac{\partial H} {\partial\bar\psi_{2n}}+
\psi_{1n}\bar\psi_{3n}\frac{\partial H} {\partial\bar\psi_{3n}}\}, \
(1\rightleftharpoons 2)\  (1\rightleftharpoons 3)
\end{eqnarray}
or for arbitrary $s$ (and ${\bf CP}^{2s}$ GCS):
\begin{equation}
i\dot\psi_{jn}=\left(1+\sum_{k=1}^{2s}|\psi_{kn}|^2\right)\left\{(1+|\psi_{jn}|^2)\frac{
\partial H}{\partial\bar\psi_{jn}}+\psi_{jn}\sum_{k=1}^{2s}\bar\psi_{jk}
\frac{\partial H} {\partial\bar\psi_{jk}}\alpha_{1k}\right\},
\end{equation}
with
\begin{eqnarray}
\alpha_{1k}=\left\{
\begin{array}{c}
0 \quad if \quad k=1 \nonumber\\
1 \quad if \quad k\not= 1, \nonumber\\
\end{array}
\right\}
\end{eqnarray}

\item {\large {Classical continuum equations of motion}}

To obtain equations of motion in the continuum approximation, we apply the
conventional procedure of expanding lattice functions, $\psi_{n\pm 1}$, in
the Taylor series up to the second order derivatives (the first nonvanishing
terms)
supposing $\lambda a_o \ll 1$:

$$\psi_{n\pm 1}=\psi (x)\pm a_o\psi'(x)+\frac{1}{2} a_o^2\psi''(x)+
O(a_o^3\psi'''),$$

where $x=a_on$.

Consider first for the sake of simplicity the ${\bf S}^2$ and ${\bf S}^{1,1}$
cases. Then the equations of motion become
\begin{equation}
i\dot\psi + \triangle\psi-2\frac{(\nabla\psi)^2\bar\psi}{1+|\psi|^2}+\delta
\frac{1-|\psi|^2}{1+|\psi|^2}=0
\end{equation}
for the ferromagnetic case and
\begin{equation}
i\dot\zeta +
\triangle\zeta+2\frac{(\nabla\zeta)^2\bar\zeta}{1-|\zeta|^2}+\delta\frac{1+
|\zeta|^2}{1-|\zeta|^2}=0
\end{equation}
for the antiferromagnetic case,
where $\Delta=\nabla^2$ is the Laplace operator.
For one-space dimensional systems these equations, being the stereographic
projections of compact and noncompact Landau-Lifshitz models defined
respectively on the sphere ${\bf S}^2$ and the hyperboloid  ${\bf S}^{1,1}$,
are gauge equivalent to various integrable versions of Nonlinear Schrodinger
Equation (NSE) [7]. In the particular case of $\delta>o$, (81) is equivalent
to the cubic attraction type NSE and (82) to the repulsive type NSE. The
latter, usually called the Ginzburg-Landau equation, describes superfluid
phenomenology and gives the correct Bogolubov excitation spectrum. What is
probably more intriguing is that the $\sigma-model$ version of (82) even
gives a correct quasiclassical description of the Bogolubov condensate,
thereby pointing out at the intimate coupling of antiferromagnetism and
superfluidity. Quite a detailed discussion of this equivalence along with
solutions of the equations can be found in [2] and [7].

Naturally, the question occurs of to what extent these continuum models can
be related to the initial lattice ones. It is easy to show that at $D\geq2$
such continuum systems lose stationary localized solutions.
In what follows, our goal will be the  ${\bf CP}^2$ model. In this concern,
as we have already seen, the most interesting is model (2) with single-ion
anisotropy. Equations of motion then assume the simplest form in terms of real
functions $(\theta, \phi, \gamma, g)$ and $D=1$:
\begin{eqnarray}
\dot\phi\ = &  & \frac{cos2g}{sin\theta}\theta_{xx} -cos2gcos\theta\phi_x^2
-4\frac{sin2g}{sin\theta}g_x\theta_x \nonumber\\  & + & \delta\frac{cos
\theta}{cos2g}(4sin2gcos2\gamma-1) \nonumber\\
\dot\theta\ = & - & cos2g(sin\theta\phi_{xx}+2cos\theta\phi_x\theta_x)+
4sin2gsin\theta g_x\theta_x \nonumber\\
 & - & \frac{\delta}{2}tan2gsin2\theta sin2\gamma \nonumber\\
\dot\gamma\ = &  & 2sin2g(g_{xx}+2cotan\theta g_x\theta_x)-cos2g(cotan\theta
\theta_{xx}-4g_x^2-\theta_x^2-\phi_x^2)- 2cos2g \nonumber\\
 & + & \delta[\frac{cos^2\theta}{cos2g}(1-sin2gcos2\gamma)+\frac{1}{2}cotang
 sin^2\theta cos2\gamma] \nonumber\\
\dot g\ = &  &  \frac{\delta}{2}sin2\gamma sin^2\theta
\end{eqnarray}

In order to make the models comparable we give the equations for the exchange
 anisotropy model (1):
\begin{eqnarray}
\dot\phi & = & \frac{cos2g}{sin\theta}\theta_{xx} -cos2gcos\theta\phi_x^2
 -4\frac{sin2g}{sin\theta}g_x\theta_x -2\delta cos2gcos\theta \nonumber\\
\dot\theta & = & -cos2g(sin\theta\phi_{xx}+2cos\theta\phi_x\theta_x)+
4sin2gsin\theta g_x\theta_x \nonumber\\
\dot\gamma & = & 2sin2g(g_{xx}+2cotan\theta g_x\theta_x)-cos2g(cotan\theta
\theta_{xx}-4g_x^2-\theta_x^2-\phi_x^2)- 2cos2g \nonumber\\
\dot g & = & 0
\end{eqnarray}

From (83) and (84), one can easily infer conclusions one and two of the
next section and that system (84) reduces to Landau-Lifshitz equations
when $g=0$.

\item {\large {Conclusion}}

Summarizing the results, we can infer the following: \\
1) Spherically symmetric systems and systems with exchange anisotropy,
regardless of its magnitude and sign, can be treated via ${\bf CP}^1$ GCS
because the Hamiltonian in this case does not contain correlators, and $ S^2$
and $q^2$ are conserved separately, $\partial_tS^2=\partial_tq^2=0$ (q is a
cyclic coordinate). Since in both ground states (or in the domain regions)
$q^2=0$, the same thing occurs in the transition (intermediate) region, the
domain wall. From the physical point of view, these nonlinear systems behave,
in a sense, as linear for they do not excite higher "harmonics", in our case,
higher moments. The Landau-Lifshitz phenomenology should work well for these
and if, nevertheless, they display certain spin reduction, it should be
attributed, in the scope of our consideration, to stochastic (chaotic)
processes. This is one of the goals in order to investigate the underlying
lattice models.\\
2) Systems with single-ion anisotropy, regardless of the anisotropy
coefficient and its sign, should be treated via ${\bf CP}^2$ GCS for $s=1$
and higher ${\bf CP}^{2s}$ for higher spins. Here pair products of the spin
operators at the same site are in the Hamiltonians and therefore
$\partial_tS^2\not=0$.  Ground states can also depend on $q$ (see (66) and
(68)) so the spin reduction can occur even in the ground states and, in
principle, may be attributed to both mechanisms: excitations of higher moments
 and stochastic processes. \\
3) The same conclusions have to take place for
antiferromagnets and SU(1,1)/U(1) GCS. \\
4) For the systems with large enough exchange anisotropy, it is necessary to
develop techniques based on the q-deformed algebras, in the sense that the
algebra deformation is related to anisotropy rate.

Along with impressive results, obtained for the models considered in the scope
of the continuum approximation, especially concerning their dynamical
properties, there is still, in fact, nearly nothing serious about the dynamics
of the corresponding lattice models. Therefore, the study of the dynamical
behaviour of such models is of a great interest from both the theoretical
point of view and physical applications. This study should consist of
analytical research as well as computer modelling.

\item {\large\bf { Acknowledgment}}

V.G.M. is grateful to Alan Bishop and Gary Doolen for valuable discussions
and a support during the first period of his stay at the Center for
Nonlinear Studies at LANL.  This work was done under auspices of the U.S.
Department of Energy and the Consejo Nacional de Ciencia y Tecnologia, 
M\'exico.


\end{enumerate}

\end{document}